# Inventions on expressing emotions In Graphical User Interface
## A TRIZ based analysis

**Umakant Mishra**

Bangalore, India

http://umakantm.blogspot.in

**Contents**

## 1. Introduction

The graphical user interface has brought great values in the field of human computer interface. However, the conventional GUI is more mechanical and does not recognize or communicate emotions. The modern GUIs are trying to infer the likely emotional state and personality of the user and communicate through a corresponding emotional state.

Emotions can be expressed in graphical icons, sounds, pictures and other means. Icons are very popularly used to do this job. An icon is a small pictorial representation of some larger set of information. Icons are designed to trigger, through visual perception, operator concepts that communicate the content or operation of the system in a quick manner.

The emotions are found to be useful in especially in communication software, interactive learning systems and other adaptive environments. The field of robotics uses emotions very deeply. Various mechanisms have been developed to express emotions through graphical user interfaces. This article will illustrate some interesting inventions selected from US patent database.

## 2. Inventions on expressing emotions

### 2.1 Method and apparatus for directing the expression of emotion for a graphical user interface (5732232)

**Background problem**

The popularity of computer and Internet has made the graphical user interface more advanced. Using emotions in graphical user interface can improve the functionality of the interface. There is a need to display emotions on the GUI represented by facial expressions.

**Solution provided by the invention**

Patent 5732232 (invented by Brush II et al., assigned by IBM, issued March 1998) provides a display mechanism for showing human emotion on a computer by showing a face.

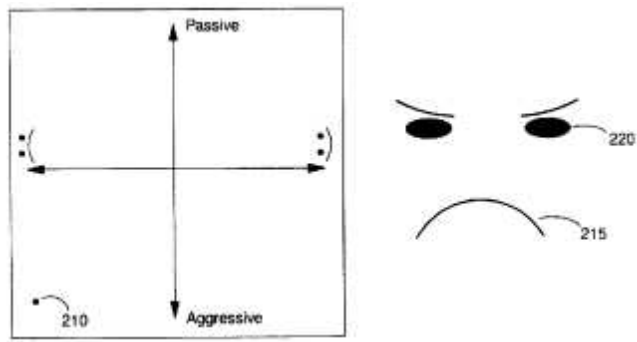

The invention shows a two-dimensional emotion graph on the screen. The user clicks on a point on the graph by using a mouse or so. When the user chooses the point on the plane the method translates the point into an appropriate value and generates a human face using the manipulation and positioning of the eyes, eyebrows and mouth.

**TRIZ based analysis**

The invention uses the picture of a human face on the graphical user interface to express emotion to the computer (Principle-28: Mechanics substitution).

The invention converts a point on the graph to a human face showing a particular emotion (Principle-36: Conversion).

### 2.2 Apparatus for interactively editing and outputting sign language information using graphical user interface (5734923)

**Background problem**

Generally a sign language is used to describe public information like how to handle an ATM used in banks, equipments in trains and flights, and other equipments in public places. Normally there has been a general method of displaying a sign language by showing different pictures and photographs. In this case one sentence of spoken language may correspond to a series of sign language images.

Besides the deaf or hearing-impaired people use a sign language and talk with the hands and fingers in their everyday life. In order for the hearing impaired people to be able to use the information output system, it is important to use a sign language in addition to the conventional expression in characters and voice. It is necessary to provide information output from a computer to allow hearing impaired handicapped people to easily understand.

**Solution provided by the invention**
The invention (Patent 5734923, invented by Sagawa et al. assigned by Hitachi Ltd, issued Mar 1998) displays information in a sign language so that the hearing impaired handicapped people can easily understand. The method displays descriptive information in combination with pictures, photographs, and video images as well as characters voice and sign-language images.

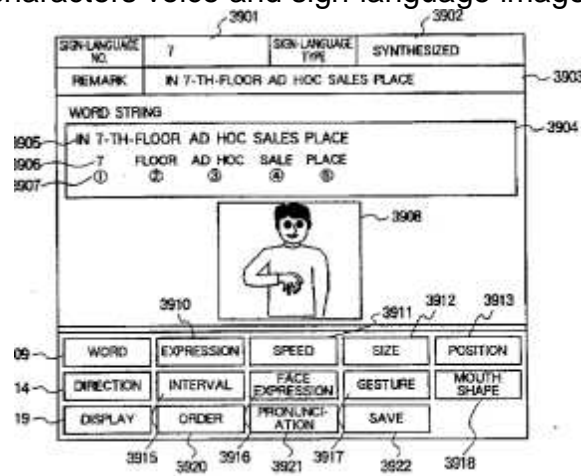

According to the invention the basic action patterns are previously stored in the computer on a word-by-word basis. When it is desired to display the sign-language animations the previously registered word images are suitably connected to make natural sign language texts.

**TRIZ based analysis**
The invention wants to supplement a sign language along with the conventional language for the convenience of hearing impaired users (Principle-8: Counterweight).
The invention uses a sign language in place of the conventional character based language (Principle-28: Mechanics substitution).

The method displays a combination of characters, photographs, pictures, voice and sign languages for better expression (Principle-40: Composite).

The method converts the natural language to a sign language using a series of hand positions and word images (Principle-36: Conversion).

## 2.3 GUI to communicate attitude or emotion to computer program (5977968)

**Background problem**

There are some computer programs, especially games, which interacts with the user depending on the feeling and mood of the user. When the mood of the user changes, the reaction of the characters of the game changes accordingly. Normally the user chooses his mood from the moodbar.

This mechanism has a limitation, although the user can express his mood through the toolbar, the game does not consider other aspect of the user's personality or emotion. There is a need to communicate these aspects of the user to the computer program.

**Solution provided by the invention**

Le Blanc invented a method of communicating attitude or emotion to a computer program (Patent 5977968, Assigned to Mindmeld Multimedia Inc., Nov 99). According to the invention, the computer provides an interactive interface for the user to communicate emotions to the computer. The expression of the face changes according to the user's movement of a cursor over the face. The user sets the expression on a face to correspond with his attitude, and the situation on the computer changes according to the attitude set by the user.

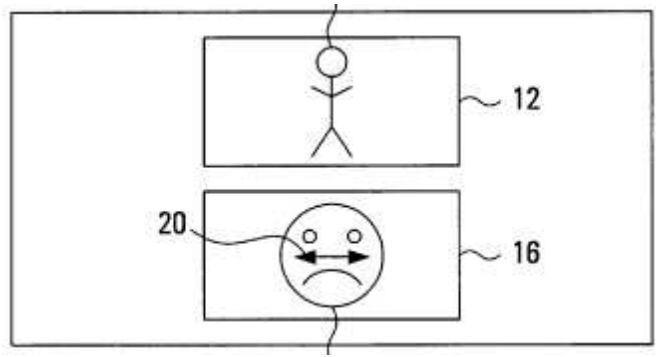

**TRIZ based analysis**

The invention communicates with the computer by expressing emotions instead of selecting menu items or command buttons (Principle-28: Mechanics substitution).

The interface is interactive where the user can change the expression by moving the cursor over the face (Principle-15: Dynamize).

## 2.4 Method and system for selecting an emotional appearance and prosody for a graphical character (6064383)

**Background problem**

Emoticons are used to express the emotion of a user during a chat session or other process of communication. For example, ":-)" or "☺" indicates happy and ":-(" or "☹" indicates sad. This system is not efficient to convey the irony or sarcasm. Moreover, it is often difficult for a user to convey the desired intensity of emotion. A character may have a single happy appearance and, thus, it may be difficult for a character to convey different degrees of happiness.

**Solution provided by the invention**

Patent 6064383 (invented by Skelly, assigned by Microsoft Corporation, issued May 2000) discloses a method of selecting a character that corresponds with an emotion and its intensity. According to the invention a user interface component has sub-regions and sub-areas, which are associated with a value for first degree and second degree of freedom. When the user positions the position indicator within a selected sub-area of one of the sub-regions, the facial expression is changed according to the value assigned to the sub-area and sub-region.

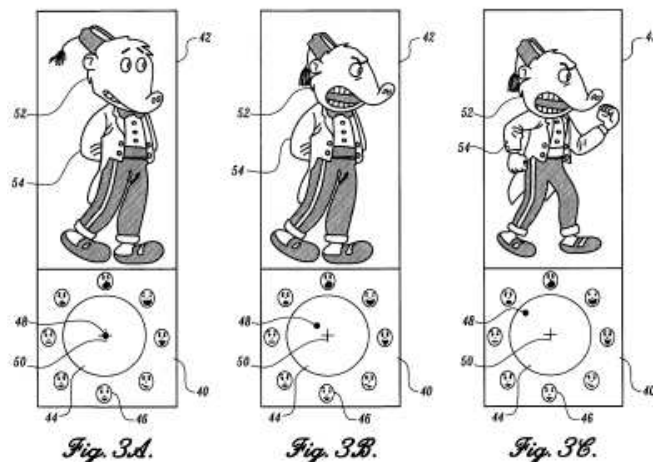

The user can manipulate the possible emotions and intensity of emotions for a character by moving the mouse inside the circular interface. The resulting character is displayed on the display device for user preview. Finally the user can choose any of the expression to use in chatting or other situations.

**TRIZ based analysis**

The user interface component is partitioned into sub-regions associated with specific emotions. Each sub-region is again divided into sub-areas, which are associated with an emotional state (Principle-1: Segmentation).

The invention generates the emotion and the intensity of the emotion (Principle-17: Another dimension).
The user can assign a specific emotion and an intensity of emotion by moving the mouse pointer inside the interface (Principle-15: Dynamize).

## 3. Other related inventions

System, method and article of manufacture for detecting emotion in voice signals through analysis of a plurality of voice signal parameters (US Patent 6151571, Invented by Pertrushin, assigned by Andersen Consulting, issued Nov 2000).

Modeling emotion and personality in a computer user interface (US Patent 6185534, invented by Breese et al., assigned by Microsoft Corporation, issued Feb 2001).

Modeling and projecting emotion and personality from a computer user interface (invented by Ball et al., assigned by Microsoft Corporation, issued Apr 2001).

System and method for a telephonic emotion detection that provides operator feedback (invented by Pertushin, assigned by Accenture LLP, issued Nov 2002)

Robot and control method for controlling the robot's emotions (invented by Osawa, assigned by Sony Corporation, issued Feb 2003)

Adaptive emotion and initiative generator for conversational systems (invented by Kleindienst et al., assigned by International Business Machines, issued Jul 2003).

Method and apparatus for analyzing affect and emotion in text (invented by Kantrowitz, assigned by Justsystem Corporation, issued Sep 2003).

Apparatus and methods for detecting emotions (invented by Liberman, assigned by Liberman, issued Oct 2003).

## 4. Summary

The concept of graphical user interface is getting into a new paradigm by including emotions. Communicating emotions can add great values to any software. At the current stage of technological development, although the use of emotions is limited only to gaming and communication software, there exists tremendous scope for using the technology in various types of software.

The inventions illustrated above propose various types of graphical icons and various methods of expressing human emotions to the computer. We can expect to see many more inventions in future on the creating, expressing and communicating emotions in graphical user interfaces.